\documentclass[%
reprint,
superscriptaddress,
amsmath,amssymb,
prd,
]{revtex4-2}

\usepackage{graphicx,wrapfig,lipsum}
\usepackage{subeqnarray}
\usepackage{url,hyperref} 
\usepackage{color}
\usepackage{ulem}
\usepackage{verbatim}

\usepackage{natbib}
\newcommand{\req}[1]{Eq.\,({\ref{#1}})}

\begin{document}
\title{{\normalsize\sc Dedicated to Professor Iwo Bia{\l}ynicki-Birula on His 90th Birthday}\\ \vspace{0.75cm} Born-Infeld nonlinear electromagnetism in relativistic heavy ion collisions}
\thanks{This article belongs to the special issue of Acta Physica Polonica A printed in honor of Professor Iwo Bia{\l}ynicki-Birula on the occasion of his 90th birthday (Ed. Tomasz Sowinski, DOI:10.12693/APhysPolA.143.S0).}
\author{Will Price}
\email{wprice@arizona.edu}
\affiliation{Department of Physics, The University of Arizona, Tucson, AZ, 85721, USA}
\author{Martin Formanek}
\email{martin.formanek@eli-beams.eu}
\affiliation{ELI Beamlines facility, The Extreme Light Infrastructure ERIC, 252 41 Doln\'{i} B\v{r}e\v{z}any, Czech Republic}
\author{Johann Rafelski}
\email{johannr@arizona.edu}
\affiliation{Department of Physics, The University of Arizona, Tucson, AZ, 85721, USA}

\date{June 13, 2023}

\begin{abstract}
We study the effect of the limiting field strength of Born-Infeld electromagnetism on the dynamics of charged particle scattering. We formulate the Born-Infeld limiting field in an invariant manner, showing that it is the electric field-dominated eigenvalue `$a$' of the field tensor $F^{\mu\nu}$ which is limited rather than the individual field vectors. Heavy ion collisions in particular provide uniquely large values of the field invariants that appear in the BI action, amplifying nonlinear effects.  Thus ``$a$'' is the dominant input into the force between heavy ions that we use to compute the scattering angle as a function of impact parameter. We evaluate the BI effects, showing relevance at small impact parameters and exhibiting their dependence on the value of the limiting field strength.
\end{abstract}

\maketitle


\section{Introduction}
In recent years, relativistic heavy ion collisions have been explored as a testing ground for strong electromagnetic (EM) field effects~\cite{Voronyuk:2011jd, Oliva:2019kin,ATLAS:2017fur,Klusek-Gawenda:2016euz,Mazurek:2021ahz,Baur:2003ar}. In peripheral collisions with large impact parameter, electromagnetic forces dominate the scattering processes, and strong field effects such as light-light scattering and spontaneous pair production can be observed. 

In this paper, we suggest that heavy ion collisions also can serve as a potential means of exploring classical, nonlinear electromagnetic effects. Specifically, we will study the nonlinear Born-Infeld (BI) theory of electromagnetism and its effect on the scattering of relativistic heavy ions. The nonlinearity in BI theory is dependent on the EM field tensor invariants, $\mathcal S$ and $\mathcal P$ (defined in Appendix~\ref{sec:appendix}), which are quite large in relativistic heavy ion collisions. 

Nonlinear EM theories inherently contain a characteristic electric field strength scale, $E_0$. In the BI theory, this field scale acts as an upper limit to the electric field in the rest frame of a particle~\cite{Born:1934gh,Bialynicki-Birula:1983}. This feature yields a finite electromagnetic mass of the particle, without the divergences that appear in Maxwell's theory, and a vanishing electromagnetic self-stress on the particle~\cite{Rafelski:1972fi}. Additionally, BI theory is the only nonlinear theory of EM in which wave velocity does not depend on its polarization~\cite{Boillat:1970gw}, and the waves can be linearly superposed with a constant background field~\cite{Bialynicka-Birula:1979ilr}. 

We begin by reviewing the Lagrangian and field equations of BI theory in Section~\ref{sec:BI_theory}. We determine the form of the Lagrangian in terms of the EM field tensor eigenvalues $a$ and $b$ (see Appendix \ref{sec:appendix}). This allows us to formulate the BI limiting field condition in an invariant manner, rather than the usual picture where $E_0$ is the limiting electric field in the rest frame of the particle. Our formulation shows that the electric-like eigenvalue $a$ is the true bounded quantity. 

In Section~\ref{sec:force}, we formulate the equations of motion of the colliding ions. In order to yield a soluble problem we approximate the ions as test particles undergoing Lorentz force motion. We then determine the EM fields of each ion separately by first solving the BI field equations in the rest frame and then Lorentz boosting to the center-of-momentum frame. This approach allows us to sidestep the difficulties that arise when one considers the force acting on a BI particle~\cite{Feenberg:1935zz, Dirac:1965, Chruscinski:1997jd, Kiessling:2003dm}, which we will turn our attention to in future work.

We present numerical results for the motion of the ions in Section~\ref{sec:results}. In particular, we look at the scattering angle dependence on impact parameter for a range of $E_0$ values. We see that the BI limiting field suppresses the scattering angle when the impact parameter is small, allowing the ions to approach each other more closely than in the framework of Maxwell electrodynamics. For larger values of $E_0$, this requires smaller impact parameters. 


\section{Born-Infeld electromagnetism} \label{sec:BI_theory}
\subsection{Review of well-known relations}
In this section we review the equations of BI theory and formulate the limiting field condition in a Lorentz invariant form. The definition of all relevant mathematical quantities and notation can be found in Appendix~\ref{sec:appendix}. The free Lagrangian of the Born-Infeld theory is, in flat spacetime with a metric $g_{\mu\nu}$, given by
\begin{equation}
    \mathcal{L} = \epsilon_0 E_0^2\left[1-\sqrt{-\text{det}\left(g_{\mu\nu}+\frac{cF_{\mu\nu}}{E_0}\right)}\right]\,,
\end{equation}
where $F_{\mu\nu}$ is the EM field tensor. Computing the determinant yields the equivalent form
\begin{equation} \label{eq:BI_lagrangian}
    \mathcal{L} = \epsilon_0 E_0^2\left(1-\sqrt{1+\frac{2\mathcal{S}c^2}{E_0^2}-\frac{\mathcal{P}^2c^4}{E_0^4}}\right) \, ,
\end{equation}
where $\mathcal S$ and $\mathcal P$ are the EM field tensor invariants: see Eqs. (\ref{eq:S}) and (\ref{eq:P}). In the limit of large $E_0$, \req{eq:BI_lagrangian} becomes the Maxwell Lagrangian, 
\begin{equation}
\mathcal L \rightarrow\mathcal S/\mu_0 \, , \, \text{as} \, \, E_0 \rightarrow \infty \, .
\end{equation}

Upon coupling~\req{eq:BI_lagrangian} to a current density $j^\mu$ and performing a variation with respect to the four-potential $A^\mu$, one arrives at the BI field equations
\begin{align} \label{eq:BI_eq1}
    \partial_\mu\mathcal{H}^{\mu\nu} &= j^\nu\,, \\ 
    \partial_\mu \widetilde{F}^{\mu\nu} &= 0 \,, \label{eq:BI_eq2} 
\end{align}
where the displacement field tensor $\mathcal{H}^{\mu\nu}$ is defined to be
\begin{equation} \label{eq:displacement_field}
    \mathcal{H}^{\mu\nu} \equiv 2\frac{\delta \mathcal{L}}{\delta F_{\mu\nu}} =\frac{1}{\mu_0}\frac{F^{\mu\nu}-\frac{\mathcal{P}c^2}{E_0^2}\widetilde{F}^{\mu\nu}}{\sqrt{1+\frac{2\mathcal{S}c^2}{E_0^2}-\frac{\mathcal{P}^2c^4}{E_0^4}}}
\end{equation}
and $\widetilde{F}^{\mu\nu}$ denotes the dual EM tensor defined in \req{eq:dualF}. The BI equations then take a form identical to the Maxwell equations in a dielectric medium. The nonlinear BI effects can then be interpreted as the effect of the dielectric medium on the field close to its source. 

One can invert~\req{eq:displacement_field} and solve for $F^{\mu\nu}$ (see Ref.~\cite{Born:1934gh}), yielding
\begin{equation} \label{eq:em_field}
    F^{\mu\nu} = \mu_0\frac{\mathcal{H}^{\mu\nu}+\frac{\mathcal{Q}\mu_0^2c^2}{E_0^2}\widetilde{\mathcal{H}}^{\mu\nu}}{\sqrt{1-\frac{2\mathcal{R}\mu_0^2c^2}{E_0^2}-\frac{\mathcal{Q}^2\mu_0^4c^4}{4E_0^4}}} \, ,
\end{equation}
where $\mathcal{R}$ and $\mathcal{Q}$ are invariants of the displacement field tensor defined in Eqs. (\ref{eq:R}) and (\ref{eq:Q}).


\subsection{Eigenvalue formulation of BI theory}

We can also formulate BI theory in terms of the EM field tensor eigenvalues, $\pm a$ and $\pm ib$, often used in the context of the Euler-Heisenberg-Schwinger effective action of quantum electrodynamics (QED)~\cite{Schwinger:1951nm} (see~\req{eq:F_eigenvalues}). This approach has the benefit of making the limiting field condition explicit in the action. The eigenvalues are related to the invariants $\mathcal{S}$ and $\mathcal{P}$ through
\begin{align}
	a &= \sqrt{-\mathcal{S} + \sqrt{\mathcal{S}^2 + \mathcal{P}^2}}\,,\label{eq:invariant_a}\\
	b &= \sqrt{\mathcal{S} + \sqrt{\mathcal{S}^2 + \mathcal{P}^2}}\,.
\end{align}
$a$ is the electrically dominated eigenvalue and $b$ is the magnetically dominated eigenvalue such that for zero magnetic field, $a=E/c$, and for zero electric field, $b=B$. Since the expressions under all four square roots are strictly positive, one can easily check that the following identities are true
\begin{align}
	\label{eq:ab_diff} a^2 - b^2 &= - 2\mathcal{S}\,,\\
	\label{eq:ab_prod} a^2b^2 &= \mathcal{P}^2\,,
\end{align}
as well as $a^2 > 0$ and $b^2 > 0$. This allows us to re-write the determinant in~\req{eq:BI_lagrangian} as the product of eigenvalues:
\begin{equation} \label{eq:BI_lagrangian2}
    \mathcal{L} = \epsilon_0 E_0^2\left(1-\sqrt{\prod_{k=1}^4\Lambda_k}\right)
\end{equation} 
where the eigenvalues $\Lambda_k$ of $\delta^\mu_\nu+\frac{F^\mu_\nu c}{E_0}$ are defined below~\req{eq:BI_eigenvalues}. Now~\req{eq:BI_lagrangian2} can be simplified to 
\begin{equation} \label{eq:BI_lagrangian3}
    \mathcal L = \epsilon_0 E_0^2 \left(1-\sqrt{\left(1-c^2\frac{a^2}{E_0^2}\right)\left(1+c^2\frac{b^2}{E_0^2}\right)}\right) \, .
\end{equation}
In fact, \req{eq:BI_lagrangian3} can also be obtained by directly substituting~\req{eq:ab_diff} and~\req{eq:ab_prod} directly into~\req{eq:BI_lagrangian}.

The expression under the square root in~\req{eq:BI_lagrangian3} must be positive in order for the Lagrangian to be real and the corresponding field equations to be real. Imposing this constraint yields an upper limit for the field strength, 
\begin{equation} \label{eq:limiting_afield}
	a < E_0/c \, ,
\end{equation}
while there is no limit on the value of $b$. Thus, it is the electric field-like invariant eigenvalue $a$ which in general is limited, not just the electric field. 

We now study the limiting field behavior of two different configurations of the invariants. When $\mathcal P = 0$, we have 
\begin{equation}
    a=\sqrt{-\mathcal S + |\mathcal S|} \, .
\end{equation}
For an electrically-dominated system, $\mathcal S < 0$ and $a = \sqrt{-2 \mathcal S}$. In this case, $\mathcal S$ is limited. For a magnetically-dominated system with $\mathcal S > 0$, we have $a = 0$ and the fields are therefore not limited.

When $\mathcal S = 0$, we have 
\begin{equation}
    a = \sqrt{|\mathcal P|} \, ,
\end{equation}
and therefore $\mathcal P$ is limited. 

\subsection{Value of the limiting field constant $E_0$}

As we will argue below, experimental data in relativistic heavy ion collisions or other strong field environments could be used to determine the BI limiting field constant $E_0$. Born and Infeld originally calculated the value of $E_0$ on the assumption that the experimentally measured mass of the electron is entirely electromagnetic, yielding a limiting field value
\begin{equation} \label{eq:BI_limit}
    E_0 = 1.18 \cdot 10^{20} \, \text{V/m} \, .
\end{equation}
However, we now know that the electron mass is made up in part of non-electromagnetic components. Therefore, the Born and Infeld value of $E_0$ cannot be exact. Instead, we must determine $E_0$ by studying charged particle dynamics experimentally. There were several previous studies of possible bounds on the BI limiting field constant with conflicting results~\cite{Rafelski:1971xw, Rafelski:1973fm, Kiessling:2003yg, Carley:2006zz, Davila:2013wba}. We hope that further studies of the effects of BI theory on particle dynamics in strong-field environments, such as in the present paper, as well as relatively recent papers on BI effects in laser-plasma acceleration~\cite{Dereli:2010yi, Burton:2010wg} can lay the groundwork for an experimental study of the value of $E_0$.

 It is also interesting to compare nonlinear classical theory with the non-linearity inherent in QED. We can compare the classical BI limiting field (at least the value calculated by Born and Infeld) to the field strength that appears in the Euler-Heisenberg-Schwinger effective Lagrangian of QED~\cite{Heisenberg:1936,Schwinger:1951nm,Evans:2023}
 \begin{equation}
     E_\text{EHS} = \frac{m_e^2c^3}{e\hbar} = 0.0112E_0 \, .
 \end{equation}
Thus, $E_\text{EHS}$ represents the scale at which quantum nonlinear effects set in, while $E_0$ corresponds to classical nonlinear effects. We see here that the quantum nonlinear scale is approximately two orders of magnitude smaller than the classical nonlinear scale. 

Additionally we can compare $E_0$ to the limiting acceleration value that appears in the Eliezer-Ford-O'Connell (EFO) radiation reaction (RR) force model~\cite{Price:2021zqq}. The EFO radiation reaction force can be written as
\begin{equation}
    \mathcal F^\mu_{\text{EFO}} = \tau_0 P^\mu_\nu \frac{d}{d\tau}(eF^{\nu\alpha}u_\alpha) \ ,
\end{equation}
which yields the following equation of motion:
\begin{equation} \label{eq:EFO}
    \left(g_{\mu\nu}-\frac{e\tau_0}{m}P_\mu^\alpha F_{\alpha \nu}\right)\dot u^\nu = \frac{e}{m}(F_{\mu\nu}+\tau_0 \dot F_{\mu\nu})u^\nu \, , 
\end{equation}
where the dot refers to the proper time derivative and $\tau_0$ is the characteristic RR time scale (for electrons $\tau_0 = 6.26\times 10^{-24}$ s). The orthogonal projection tensor is defined as
\begin{equation}
    P_{\mu\nu} = g_{\mu\nu} - \frac{u_\mu u_\nu}{c^2} \,.
\end{equation}
To solve for the acceleration in~\req{eq:EFO}, one must invert the tensor $g_{\mu\nu} - \frac{e\tau_0}{m} P_\mu^\alpha F_{\alpha\nu}$ by taking its determinant, which takes a similar form to the BI Lagrangian~\req{eq:BI_lagrangian}: 
\begin{equation}
    \begin{split}
    -\text{det}&\left(g_{\mu\nu} - \frac{e\tau_0}{m} P_\mu^\alpha F_{\alpha\nu}\right) \\
    &= 1+ \frac{e^2\tau_0^2}{m^2}\left(2\mathcal S + uFFu/c^2\right) \,.
    \end{split}
\end{equation}
As we have shown in Ref.~\cite{Price:2021zqq}, in certain field configurations~\req{eq:EFO} leads to an upper limit on the acceleration analogous to the BI limiting field
\begin{equation}
    a_{\text{EFO}} = \frac{c}{\tau_0} = \frac{3}{2}\frac{4\pi \epsilon_0 mc^2}{e^2} = 4.80\times 10^{31}\, \text{m/s}^2\,, 
\end{equation}
where the value is given for electrons. The electric field corresponding to this limiting acceleration is 
\begin{equation}
    E_{\text{EFO}} = E_{\text{EHS}}/\alpha = 1.53 E_0\, ,
\end{equation}
where $\alpha \approx 1/137$ is the fine-structure constant. We see that the classical limiting field appearing in the radiation reaction force is of the same order of magnitude as the BI limiting field, $E_0$, which is obtained by requiring all of the electron mass to be of electromagnetic origin.


\section{heavy ion scattering} \label{sec:force}
\subsection{Force on a BI particle}
We'll consider the scattering of two identical heavy ions, both of charge $Ze$ and mass $m$. We will approximate the ions as test particles. In this case, the force on each charge is given by the Lorentz force. For ions located at positions $\pmb{x}_1$ and $\pmb{x}_2$ and with fields $F^{\mu\nu}_1$ and $F^{\mu\nu}_2$, our system of equations is
\begin{align} \label{eq:lorentzforce1}
    m\dot u^\mu_1 &= ZeF^{\mu\nu}_2 u_{1\nu}\,, \\
    m\dot u^\mu_2 &= ZeF^{\mu\nu}_1 u_{2\nu} \, . \label{eq:lorentzforce2}
\end{align}
For symmetric collisions, the ions will have the same proper time, $\tau$. In the test particle approximation, the field of one ion acts on the other ion. The BI effects in our equations of motion come from $F^{\mu\nu}$, which is a solution of the BI field equations and therefore is restricted by the limiting field condition $a<E_0/c$. We neglect backreaction from the self-field of each ion. 

 Our test particle approach allows us to formulate a tractable problem, since a closed form expression for the force between two relativistic charges in BI theory is not known. The Lorentz force, which we use here, is the leading order force on a BI charge for small acceleration~\cite{Pryce:1936uap}. First order BI corrections to the Coulomb force between two static charges have been calculated in~\cite{Ferraro:2006rq}. 

Our first step, then, is to solve the BI equations for a single ion undergoing relativistic motion. The analogous problem for Maxwell's equations is solved by the Lienard-Wiechert fields~\cite{Jackson:1998nia}. However, in the nonlinear BI theory, the exact solution to this problem is not yet known.


\subsection{Fields of a relativistic BI particle}

To obtain an approximate solution for the field of the ion, we will assume that its acceleration is relatively small so that its rest frame is approximately inertial. In the rest frame, assuming that the ion is located at a position $\pmb{x}'_0$ with charge $Ze$, the BI equations simplify to
\begin{equation} \label{eq:coulomb}
    \nabla \cdot \pmb{D}'(\pmb x') = Ze\delta^3(\pmb{X}') \, , \, \,\pmb X' \equiv \pmb x' - \pmb x'
    _0 \, ,
\end{equation}
while the magnetic field vanishes
\begin{equation} \label{eq:magnetic_sol}
    \pmb{H}'(\pmb x') = 0 \, .
\end{equation}
Here \req{eq:coulomb} is solved by the Coulomb field
\begin{equation} \label{eq:coulomb_sol}
    \pmb{D}'(\pmb{x}') = \frac{Ze}{4\pi}\frac{\pmb{X}'}{|\pmb{X}'|^3} \, .
\end{equation}

We can then Lorentz boost the rest frame fields,~\req{eq:magnetic_sol} and~\req{eq:coulomb_sol}, to the center-of-momentum (CM) frame where the ion is moving with velocity $\pmb{v}$ with negligible acceleration. The $\pmb D$ and $\pmb H$ fields transform identically to $\pmb E$ and $\pmb B$, whose transformation formulae are given in~\cite{Jackson:1998nia}. The details of the Lorentz transformation are presented in Appendix~\ref{appendix:fields}. The displacement fields in the CM frame are given in~\req{eq:boosted_Dfield} and~\req{eq:boosted_Hfield}
\begin{align}
    \pmb{D}(t,\pmb{x}) &= \frac{Ze}{4\pi} \frac{\gamma}{R^{3}}\pmb{X} \label{eq:Dfield}\,, \\
    \pmb{H}(t,\pmb{x}) &= \frac{Ze}{4\pi}\frac{\gamma}{R^3}\pmb{v}\times \pmb{X}\, , \label{eq:Hfield}
\end{align}
where
\begin{equation}
    R = \sqrt{\pmb{X}_\perp^2 + \gamma^2\pmb{X}^2_\parallel} \, .
\end{equation}

We then compute the invariants $\mathcal R$ and $\mathcal Q$, using~\req{eq:Dfield} and~\req{eq:Hfield}, which leads to 
\begin{align}
    \mathcal R &= -\frac{1}{2}\left(\frac{Zec}{4\pi R^2}\right)^2 \label{eq:invariant_R}\,,\\
    \label{eq:invariant_Q} \mathcal Q &= 0 \, .
\end{align}
Now \req{eq:invariant_R} is obtained by noting that 
\begin{equation}
    \left(\frac{\pmb v}{c} \times \pmb X\right)^2 - \pmb X^2 = R^2/\gamma^2 \, . 
\end{equation}

We can now compute the $\pmb E$ and $\pmb B$ fields corresponding to~\req{eq:Dfield} and~\req{eq:Hfield}. In three-vector form, ~\req{eq:em_field} reads
\begin{align} \label{eq:Efield1}
    \pmb{E} &= \frac{1}{\epsilon_0}\frac{\pmb D}{\sqrt{1-\frac{2\mathcal{R}\mu_0^2 c^2}{E_0^2}}}\,,\\
    \pmb{B} &= \mu_0\frac{\pmb H }{\sqrt{1-\frac{2\mathcal{R}\mu_0^2c^2}{E_0^2}}} \label{eq:Bfield1} \, ,
\end{align}
where we have substituted $\mathcal Q = 0$. Upon combining~\req{eq:Efield1} and~\req{eq:Bfield1} with the expressions for $\pmb D$, $\pmb H$, and $\mathcal R$, we find
\begin{align}
    \pmb E &= \frac{Ze\gamma}{4\pi\epsilon_0R^3}\frac{\pmb{X}}{\sqrt{1+\left(\frac{Ze}{4\pi\epsilon_0R^2}/E_0\right)^2}} \label{eq:Efield}\,,\\ 
    \pmb B &= \frac{Ze\gamma}{4\pi c^2\epsilon_0R^3}\frac{\pmb{v}\times \pmb{X}}{\sqrt{1+\left(\frac{Ze}{4\pi\epsilon_0R^2}/E_0\right)^2}} \label{eq:Bfield} \, .
\end{align}
The behavior of these fields in the ultrarelativistic limit is studied in~\cite{Bicak:2006by}.

As a final step, we will check that our fields obey the limiting field constraint from~\req{eq:limiting_afield}. For this $\mathcal S$ and $\mathcal P$ need to be computed. By the well-known duality symmetry of BI theory~\cite{Bialynicki-Birula:1983}, 
\begin{equation}
    \mathcal P = \mu_0^2 \mathcal Q \, ,
\end{equation}
and therefore $\mathcal P$ vanishes. Computing $\mathcal S$ using~\req{eq:em_field}, we find
\begin{equation}
    \mathcal S = \mu_0^2\frac{\mathcal{R}}{1-\frac{2\mathcal{R}\mu_0^2 c^2}{E_0^2}-\frac{\mathcal Q^2\mu_0^4c^4}{E_0^4}} \, ,
\end{equation}
which simplifies to
\begin{equation}
    \mathcal S = -\frac{1}{2}\frac{\left(\frac{Ze}{4\pi\epsilon_0 R^2}\right)^2}{1+\frac{\left(\frac{Ze}{4\pi\epsilon_0R^2}\right)^2}{E_0^2}} \, .
\end{equation}
In this case $\mathcal P=0$, so we have from~\req{eq:invariant_a} $a=\sqrt{-2\mathcal S}$ and therefore
\begin{equation}
    a = \frac{\frac{Ze}{4\pi c\epsilon_0 R^2}}{\sqrt{1+\frac{\left(\frac{Ze}{4\pi\epsilon_0R^2}\right)^2}{E_0^2}}} \, .
\end{equation}
In the limit $R\rightarrow 0$, 
\begin{equation}
    a \rightarrow E_0/c
\end{equation}
and we see that our fields obey the invariant limiting field condition.


\section{Numerical results} \label{sec:results}
\begin{figure}
\includegraphics[width=\linewidth]{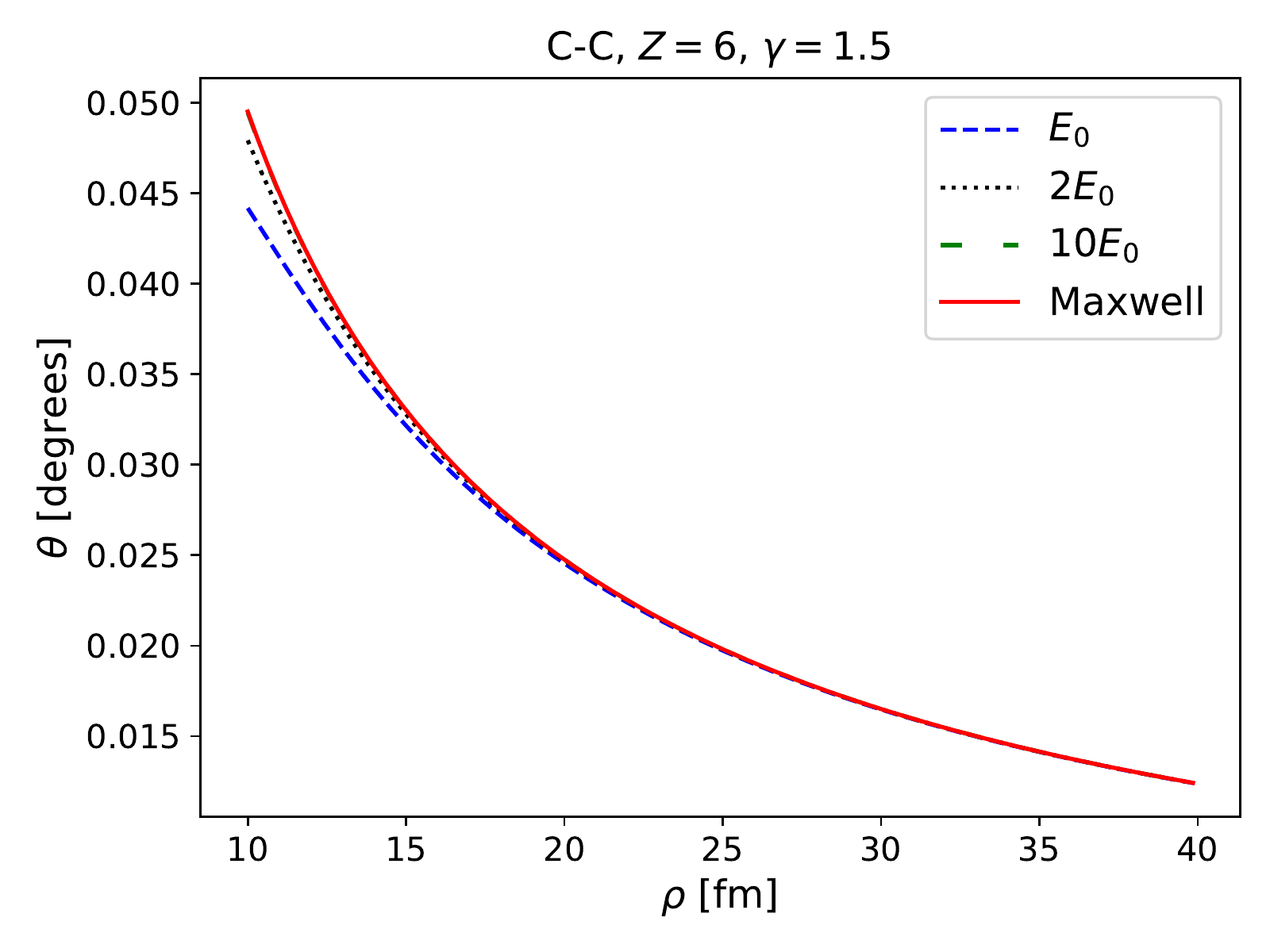}
\includegraphics[width=\linewidth]{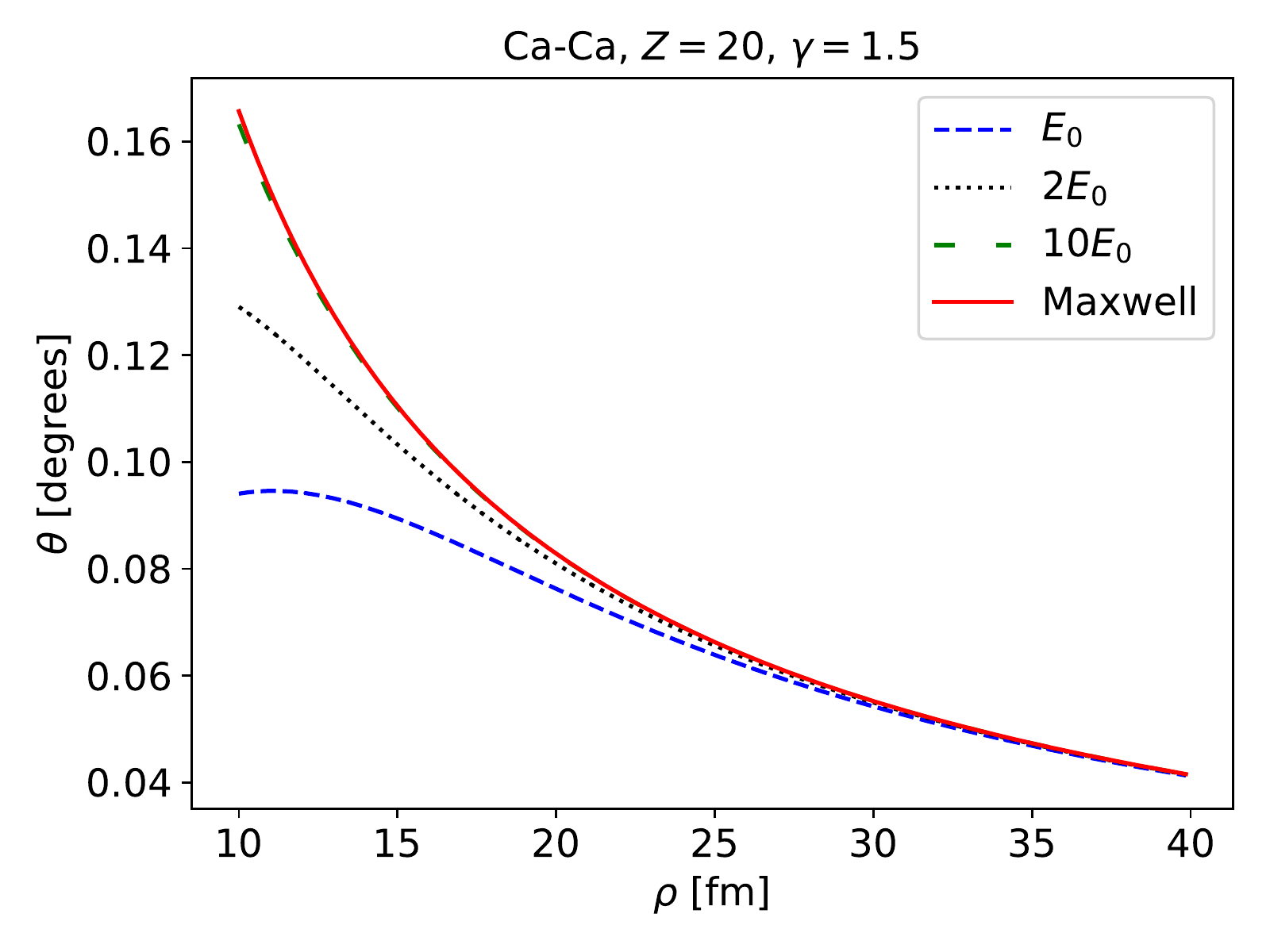}
\caption{\label{fig:angle1}Scattering angle as a function of impact parameter $\rho$ for C-C collisions (top panel) and Ca-Ca (bottom panel) and with $\gamma=1.5$.}
\end{figure}

\begin{figure}
\includegraphics[width=\linewidth]{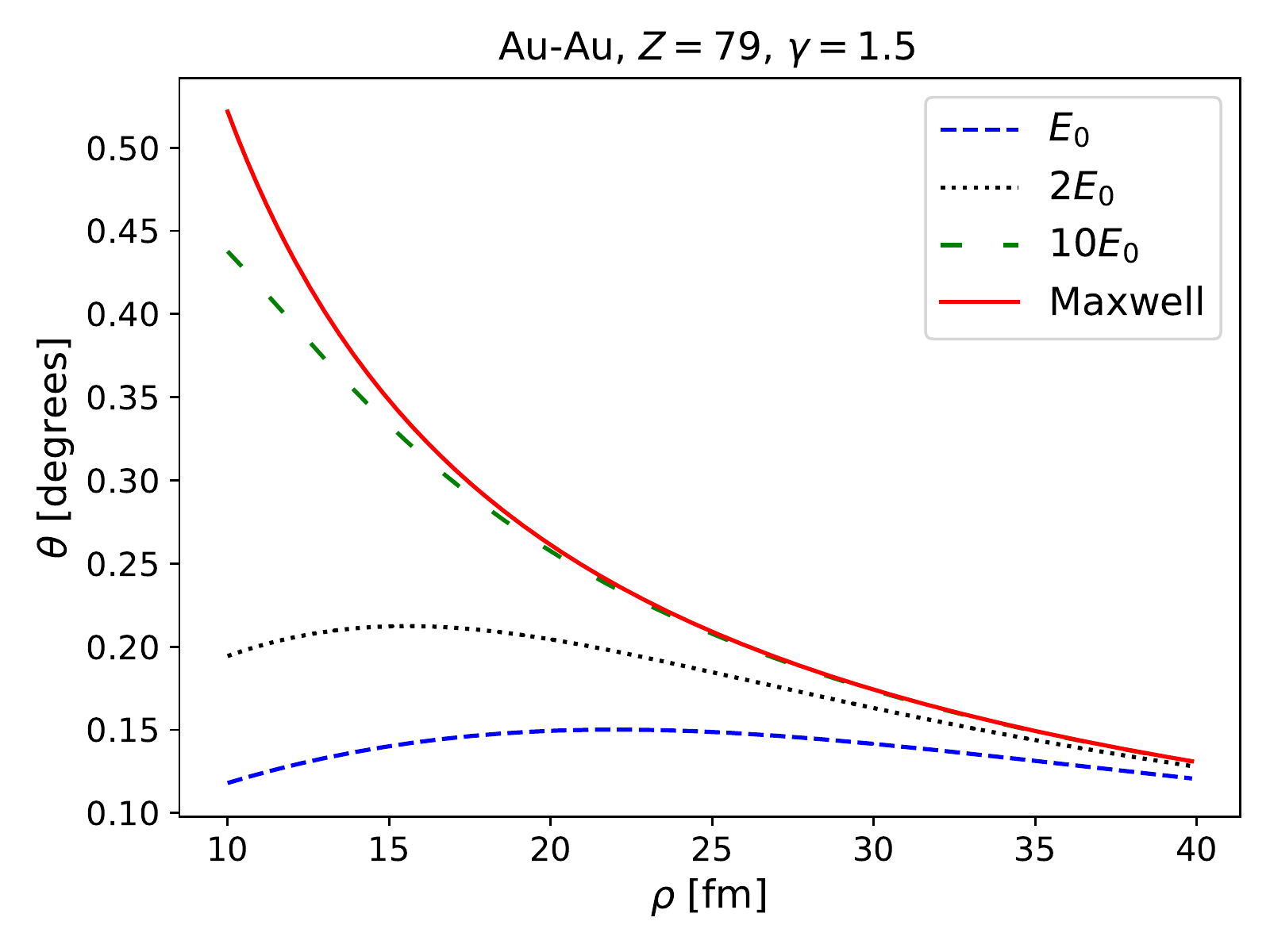}
\includegraphics[width=\linewidth]{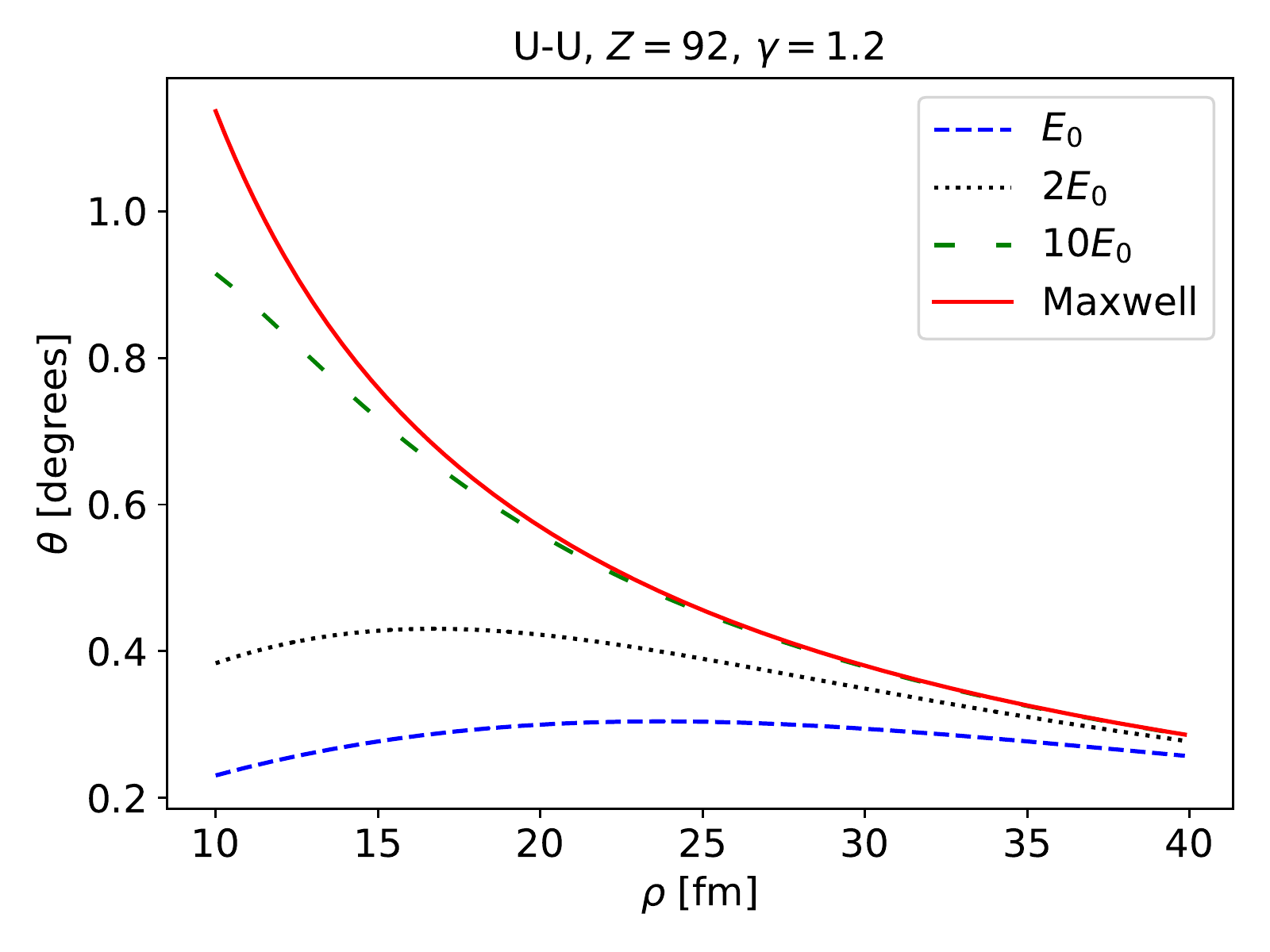}
\caption{\label{fig:angle2}Scattering angle $\theta$ as a function of impact parameter $\rho$ for Au-Au collisions with $\gamma=1.5$ (top panel) and U-U collisions with $\gamma=1.2$ (bottom panel).}
\end{figure}

Our heavy ion collisions are described by the system of equations~\req{eq:lorentzforce1} and~\req{eq:lorentzforce2} which depend on the electric and magnetic fields~\req{eq:Efield} and~\req{eq:Bfield}. We integrate the equations using the fifth-order Runge-Kutta-Dormand-Prince method~\cite{Dormand:1980} and use the constraint $u^2 = c^2$ as an estimation of numerical error. The ions are assumed to start from $\tau=-\infty$ with incoming velocities $\pm v_0$, as well as transverse separation in the $y$-direction by an impact parameter $\rho$. The initial longitudinal separation is taken to be large enough such that the results are converged with respect to its small variations. We use $\rho$ for impact parameter rather than the usual $b$ since the latter is reserved for the EM tensor eigenvalue in this paper.  

We present results for four pairs of colliding ions: Au-Au, U-U, Ca-Ca, and C-C to study a range of charge-to-mass ratios. We compute the scattering angle $\theta$ by comparing incoming and outgoing momentum vectors for the ions. 

In Figure~\ref{fig:angle1}, we plot the relationship between $\theta$ and $\rho$ for C-C (top panel) and Ca-Ca (bottom panel) at $\gamma=1.5$. In Figure~\ref{fig:angle2}, we present the scattering angle for Au-Au at $\gamma=1.5$ (top panel) and U-U at $\gamma=1.2$ (bottom panel). The scattering angle is computed for the Born and Infeld value of $E_0$ given in~\req{eq:BI_limit} as well as for $2E_0$, $10E_0$, and the Maxwell theory corresponding to $E_0\rightarrow \infty$. The results are shown for $\rho>10$ fm to ensure that the collisions remain peripheral and the point particle approximation remains valid.

Our results show that BI effects are most significant at smaller impact parameters, where the field seen by each ion is bounded. This is in contrast to Maxwell theory, where the force approaches infinity for smaller and smaller impact parameters. The limiting BI field significantly suppresses the scattering angle at lower impact parameters, with scattering angle decreasing moreso as $E_0$ decreases. At larger impact parameters, BI effects are relevant for smaller values of $E_0$ and become negligible at $10E_0$ and above. 

BI effects are additionally amplified for large $Z$. The C-C results in bottom panel of Figure~\ref{fig:angle1} show a relatively small BI contribution to the scattering, even at $1E_0$. In contrast, the Au-Au collisions in top panel of Figure~\ref{fig:angle2} with a charge approximately 13 times larger show a significant BI effect on the scattering angle.

\begin{figure}
    \includegraphics[width=\linewidth]{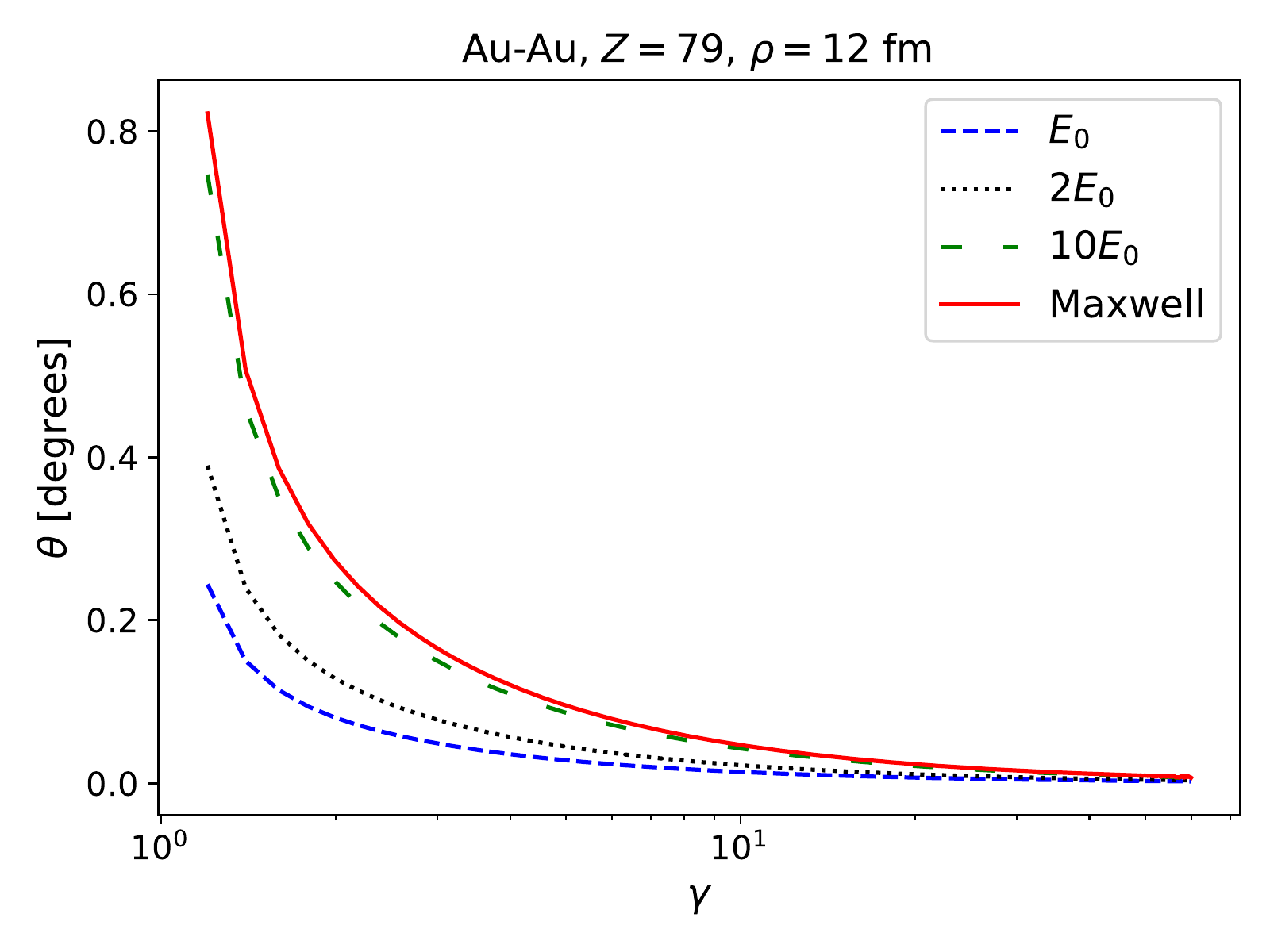}
    \caption{Scattering angle as a function of incoming $\gamma$ at a fixed impact parameter, $\rho=12$ fm.}
    \label{fig:gamma}
\end{figure}

Figure~\ref{fig:gamma} plots the scattering angle as a function of incoming $\gamma$ at a fixed impact parameter $\rho=12$ fm. The scattering angle decreases as $\gamma$, and thus the inertia of the ions $\gamma m$, increases. Due to this, BI effects are amplified at lower $\gamma$, where the scattering angle is in general larger for all values of $E_0$. The U-U plot shown in the bottom panel of Figure~\ref{fig:angle2} demonstrates the effect of large $Z=92$ as well as a smaller $\gamma=1.2$, maximizing BI effects.

\section{Conclusions}
In this work, we have studied the limiting field feature of BI theory and established that the electric-like eigenvalue of the EM field tensor $a$ is limited in all inertial frames by the electric field strength constant $E_0$. In the rest frame of the particle this becomes a limit on the electrostatic field, yielding a finite electric field energy as usual. 

We then applied BI theory to study the dynamics of relativistic heavy ion collisions. We showed that the limiting field can have a significant impact on the scattering angle at low impact parameter, when the ions pass through the region of field greatly altered by nonlinear effects. This causes a reduction in force and thus scattering angle at low impact parameter when compared to Maxwell theory. The difference between BI and Maxwell theory is most relevant for smaller values of the limiting field $E_0$ and diminishes as this limiting value is increased. 

Looking at a fixed scattering angle in Figure~\ref{fig:angle2}, say $0.4^\circ$, we see that the BI force predicts a smaller corresponding impact parameter by approximately $25$\% for $10E_0$. The BI impact parameter is approximately $\rho_\text{BI}\approx 12$ fm, while the Maxwell impact parameter is approximately $\rho_\text{Max}\approx 15$ fm. This impact parameter shift is even larger for smaller values of the limiting field constant. A shift in perceived impact parameter such as we see here would have a significant effect on all impact parameter-sensitive calculations using peripheral heavy ion collision data, such as nuclear size~\cite{Zhao:2022dac, ALICE:2022}.

Moreover, self-field effects will likely be relevant at these small impact parameters. In BI theory this includes two phenomena, i.e., radiation reaction, and the nonlinear superposition of the self-field and the external field. In future work, we will turn our attention to formulating a framework for BI particle motion that includes these effects in a self-consistent manner.

\appendix
\section{Mathematical identities and notation} \label{sec:appendix}
The conventions and notation used in this paper are as follows. We use a flat spacetime metric with a negative signature:
\begin{equation}
    g_{\mu\nu} = \text{diag}(1,-1,-1,-1) \, .
\end{equation}
The electromagnetic field tensor is given in terms of the EM four-potential $A^\mu = (\phi/c,\pmb{A})$ as
\begin{equation}
    F^{\mu\nu} = \partial^\mu A^\nu - \partial^\nu A^\mu \, ,
\end{equation}
and can be written in terms of the electric and magnetic fields $\pmb{E}$ and $\pmb{B}$ in cartesian coodinates as the following 4x4 matrix
\begin{equation}
    F^{\mu\nu} = \left(\begin{matrix}
    0 & -E_x/c & -E_y/c & -E_z/c \\
    E_x/c & 0 & -B_z & B_y \\
    E_y/c & B_z & 0 & -B_x \\
    E_z/c & -B_y & B_x & 0 \end{matrix}\right) \label{eq:F_matrix}\, .
\end{equation}
We also use the displacement field tensor $\mathcal{H}^{\mu\nu}$ which can be written in terms of the displacement fields $\pmb{D}$ and $\pmb{H}$ as 
\begin{equation}
    \mathcal{H}^{\mu\nu} = \left(\begin{matrix}
    0 & -cD_x & -cD_y & -cD_z \\
    cD_x & 0 & -H_z & H_y \\
    cD_y & H_z & 0 & -H_x \\
    cD_z & -H_y & H_x & 0 \end{matrix}\right) \, .
\end{equation}
We can also construct corresponding dual tensors
\begin{align}
    \label{eq:dualF}\widetilde{F}_{\alpha\beta} &= \frac{1}{2}\epsilon_{\alpha\beta\mu\nu}F^{\mu\nu}\,, \\
    \widetilde{\mathcal{H}}_{\alpha\beta} &= \frac{1}{2}\epsilon_{\alpha\beta\mu\nu}\widetilde{\mathcal{H}}^{\mu\nu} \, ,
\end{align}
where the totally  antisymmetric Levi-Civita symbol is defined such that
\begin{equation}
    \epsilon_{0123}=-\epsilon^{0123} = 1 \, .
\end{equation}
The dual tensors written in matrix form are
\begin{align}
    \widetilde F_{\mu\nu} &= \left(\begin{matrix}
    0 & -B_x & -B_y & -B_z \\
    B_x & 0 & -E_z/c & E_y/c \\
    B_y & E_z/c & 0 & -E_x/c \\
    B_z & -E_y/c & E_x/c & 0 \end{matrix}\right)\,, \\
    \widetilde{\mathcal H}_{\mu\nu} &= \left(\begin{matrix}
    0 & -H_x & -H_y & -H_z \\
    H_x & 0 & -D_zc & D_yc \\
    H_y & D_zc & 0 & -D_xc \\
    H_z & -D_yc & D_xc & 0 \end{matrix}\right) \, .
\end{align}
We can construct the following four field invariants from the field tensors:
\begin{align}
    \label{eq:S}\mathcal{S} &= \frac{1}{4}F^{\mu\nu}F_{\mu\nu} = \frac{1}{2}(B^2-E^2/c^2)\,, \\
    \label{eq:P}\mathcal{P} &= \frac{1}{4}F^{\mu\nu}\widetilde{F}_{\mu\nu} = \pmb{B}\cdot\pmb{E}/c\,, \\
    \label{eq:R}\mathcal{R} &= \frac{1}{4}\mathcal{H}^{\mu\nu}\mathcal{H}_{\mu\nu} = \frac{1}{2}(H^2-c^2D^2)\,, \\
    \label{eq:Q}\mathcal{Q} &= \frac{1}{4}\mathcal{H}^{\mu\nu}\widetilde{\mathcal{H}}_{\mu\nu} = \pmb{H}\cdot c\pmb{D}\,.
\end{align}

Evaluating the BI action requires a computation of the determinant of the tensor $g_{\mu\nu}+\frac{F_{\mu\nu}c}{E_0}$. The four eigenvalues of $F^\mu_\nu$ are $\lambda_k \equiv \{\pm a,\pm ib\}$, defined by~\cite{Schwinger:1951nm}
\begin{align}
	a &= \sqrt{-\mathcal{S} + \sqrt{\mathcal{S}^2 + \mathcal{P}^2}}\,,\label{eq:invariant_a}\\
	b &= \sqrt{\mathcal{S} + \sqrt{\mathcal{S}^2 + \mathcal{P}^2}}\,. \label{eq:invariant_b}
\end{align}
The parameter $a$ is electrically dominated such that for zero magnetic field 
\begin{equation}
    a = E/c \, ,
\end{equation}
and the parameter $b$ is magnetically dominated such that for zero electric field
\begin{equation}
    b = B \, .
\end{equation}
For $E/c$ and $B$ of equal magnitude we have $\mathcal S = 0$ and
\begin{equation}
    a = b = |\mathcal P| \, .
\end{equation}
In general the determinant of $F_{\mu\nu}$ can be written as the product of eigenvalues
\begin{equation}
    \begin{split} \label{eq:F_eigenvalues}
    -\text{det}(F_{\mu\nu}) &= -\det(g_{\mu\alpha}F^{\alpha}_\nu) = - \det(g_{\mu\alpha})\det(F^\alpha_\nu)\\
    &=\det(F^\alpha_\nu) = \prod_{k=1}^4 \lambda_k = -a^2 b^2 = -\mathcal P^2 \,,
    \end{split}
\end{equation}
because of the identity $\det(AB) = \det(A)\det(B)$ and in the flat spacetime background $\det(g_{\mu\nu}) = -1$. The last equality follows either from direct computation with the matrix form~\req{eq:F_matrix} or manipulation of~\req{eq:invariant_a} and~\req{eq:invariant_b}. Following the same steps, we find the determinant of $g_{\mu\nu} + \frac{F_{\mu\nu}c}{E_0}$ to be
\begin{equation} \label{eq:BI_eigenvalues}
    \begin{split}
    -\text{det}&\left(g_{\mu\nu} + \frac{F_{\mu\nu}c}{E_0}\right) = \text{det}\left(\delta^\mu_\nu + \frac{F^\mu_\nu c}{E_0}\right) = \prod_{k=1}^4 \left(1 + \lambda_k \frac{c}{E_0}\right)\\
    &= \left(1 - a \frac{c}{E_0}\right)\left(1 + a \frac{c}{E_0}\right)\left(1 - ib \frac{c}{E_0}\right) \left(1 + ib \frac{c}{E_0}\right)\\
    &= \left(1-\frac{c^2a^2}{E_0^2}\right)\left(1+\frac{c^2b^2}{E_0^2}\right) \\
    &= 1 + \frac{2\mathcal S c^2}{E_0^2} - \frac{\mathcal P^2 c^4}{E_0^4} \,,
    \end{split}
\end{equation}
which is used in the main text of this paper where $\Lambda_k \equiv 1 + \lambda_k c/E_0$ are eigenvalues of $\delta^\mu_\nu + F^\mu_\nu c /E_0$. 

\section{Fields of a uniformly moving charge} \label{appendix:fields}
The displacement fields of the ion in the CM frame are given by the Lorentz transformation of the rest frame fields~\req{eq:coulomb_sol} and~\req{eq:magnetic_sol}. The transformation equations are~\cite{Jackson:1998nia}
\begin{align}
    \pmb D_\parallel &= \pmb D_\parallel' \, ,\\
    \pmb H_\parallel &= \pmb H_\parallel' \, , \\
    \pmb D_\perp &= \gamma\left(\pmb D_\perp' - \frac{\pmb v}{c^2}\pmb \times H'\right) \, ,\\
    \pmb H_\perp &= \gamma\left(\pmb H_\perp' + \pmb v \times \pmb D'\right) \, .
\end{align}

The unprimed fields represent the fields in the CM frame, while the primes refer to quantities in the rest frame. We apply the Lorentz boost along the $-\pmb X_\parallel$ direction so that the charge is seen moving along the $+\pmb X_\parallel$ direction. Therefore, $\pmb{X}_\perp$ and $\pmb{X}_\parallel$ represent the transverse and parallel components of $\pmb{X}$ with respect to the ion velocity, $\pmb{v}$. 

The coordinate transformation is 
\begin{align}
    X^0 &= \gamma \pmb v \cdot \pmb X'/c \, ,\\
    \pmb X_\parallel &= \gamma \pmb X_\parallel' \, ,\\
    \pmb X_\perp &= \pmb X_\perp' \, .
\end{align}
The $t'$ terms that usually appear in the Lorentz transformation are absent above because the relative position vector $\pmb X' = \pmb x' - \pmb x_0'$ is taken to be the difference of two vectors at equal time in the rest frame. $X^0$ is the difference in time between observation point $\pmb x$ and source point $\pmb x_0$ in the CM frame.

In the CM frame, we then find our boosted fields to be
\begin{align}
    \label{eq:boosted_Dfield} \pmb{D}(t,\pmb{x}) &= \frac{Ze}{4\pi} \frac{\gamma}{R^{3}}\pmb{X} \,, \\
    \label{eq:boosted_Hfield}\pmb{H}(t,\pmb{x}) &= \frac{Ze}{4\pi}\frac{\gamma}{R^3}\pmb{v}\times \pmb{X}\, ,
\end{align}
where
\begin{equation}
    R = \sqrt{\pmb{X}_\perp^2 + \gamma^2\pmb{X}^2_\parallel} \, .
\end{equation}
 
 Note that all components of the fields~\req{eq:boosted_Dfield} and~\req{eq:boosted_Hfield} obtain a factor of $\gamma$ under the transformation. The parallel components acquire a factor of $\gamma$ from the coordinate transformation, while the $\gamma$ in the transverse components comes from the field transformation.

To validate our results for the $\pmb D$ and $\pmb H$ fields of a charge in uniform motion, obtained by Lorentz transformation, we show that~\req{eq:boosted_Dfield} and~\req{eq:boosted_Hfield} are consistent with the Lienard-Wiechert solution of Maxwell's equations. Starting with the Lienard-Wiechert solution, it can be shown that the electric field of a charge in uniform motion can be written as (see Ref.~\cite{Melia:2001}, page 105-106, Example 4.5)
\begin{equation} \label{eq:LW_Efield}
    \pmb E = \frac{Ze}{4\pi\epsilon_0\gamma^2} \frac{\pmb X}{|\pmb X|^3\left(1-\frac{v^2}{c^2} \sin^2\psi\right)^{3/2}} \, ,
\end{equation}
where $\psi$ is defined as the angle between the position vector $\pmb X$ and the direction of motion:
\begin{equation}
    |\pmb X_\perp| = |\pmb X| \sin \psi  \, .
\end{equation}
The denominator in~\req{eq:LW_Efield} can be re-written as
\begin{equation}
\begin{split}
    |\pmb X|^3 \left(1-\frac{v^2}{c^2} \sin^2 \psi\right)^{3/2} &= |\pmb X|^3 \frac{\left(|\pmb X|^2 -\frac{v^2}{c^2}\pmb X_\perp^2 \right)^{3/2}}{|\pmb X|^3} \\
    &= \left( \pmb X_\parallel^2 + \pmb X_\perp^2/\gamma^2\right)^{3/2} \\
    &= R^3/\gamma^3 \, .
\end{split}
\end{equation}
Then \req{eq:LW_Efield} becomes
\begin{equation}
    \pmb E = \frac{Ze}{4\pi\epsilon_0}\frac{\gamma}{R^3}\pmb X \, ,
\end{equation}
which is consistent with~\req{eq:Dfield}. The magnetic field is obtained by $\pmb B = \pmb v \times \pmb E/c^2$, which is consistent with~\req{eq:Hfield}.

\vskip 0.5cm

\end{document}